\documentstyle[12pt]{article}
\begin{document}

\title{\bf Comments on the Unification of  Electromagnetism and Gravitation
through ``Generalized Einstein manifolds"}
\author{
{\bf M. Panahi}$\dagger$ , \ {\bf M. Mehrafarin}$\dagger \ddagger$
\\ \normalsize $\dagger$ Centre for Theoretical Physics and Mathematics\\
\normalsize Atomic Energy Organisation of Iran, Tehran, Iran\\
\normalsize MPanahi@vax.ipm.ac.ir\\
{\normalsize and}\\
\normalsize $\ddagger$ Department of Physics,
Amir Kabir University of Technology, Tehran, Iran\\
\normalsize Mehrafar@cic.aku.ac.ir}

\date{}
\maketitle

\begin{abstract}
\noindent
H. Akbar-Zadeh has recently  proposed (J Geom Phys 17 (1995)
342) a new geometric formulation of Einstein--Maxwell
system {\it with source} in terms of what are called
``Generalized Einstein manifolds". We show that, contrary to the claim,
Maxwell equations have not been derived in this formulation and that, the
assumed equations can be identified only as {\it source-free}
Maxwell equations in
the proposed geometric set up.
A genuine derivation of source-free Maxwell equations
is presented within the same framework.
We draw a conclusion that the proposed unification scheme can pertain only to
source-free situations. \\
\end{abstract}

\begin{description}
\item[\it  Keywords:]
Generalized Einstein manifolds; Finsler Geometry;
Einstein\mbox{--}Maxwell system
\item[\it  1991 MSC:]
83Exx, 83C22
\item[\it  PACS:]
 04.40.Nr, 04.50.+h
\end{description}

\vspace{1.0cm}
\noindent In a recent article \cite{H1}, using the tangent bundle approach to
Finsler Geometry, H. Akbar-Zadeh  has introduced a class of Finslerian
manifolds called ``Generalized Einstein manifolds'.
These manifolds are obtained through some constrained metric variations
on an action functional depending on the curvature tensors.
The author has then
 proposed a new scheme
for the unification of electromagnetism and gravitation, in which the
spacetime
manifold, $M$, with its usual pseudo-Riemannian metric,   $g_{ij}(x)$,
is endowed with a Finslerian
connection containing the Maxwell tensor, $F^{ij}(x)$. Following this sheme,
the author arrives at a
class of Generalized Einstein manifolds
containing the solutions of Einstein\mbox{--}Maxwell
equations. As for Maxwell equatins,
they are  decrared \cite[pp 343 and 378]{H1}
to have been
obtained by means of Bianchi
identities.
We wish to point out the following
flaws in the treatment of Einstein\mbox{--}Maxwell system.

First consider the treatment of Maxwell equations.   Through some constrained
metric variarions, and the use of
Bianchi identities, the author arrives at  \cite[eq (5.55)]{H1}:
            \begin{equation}
            \nabla _i F^{ij} = \mu _1 \, u^j \, ,
            \end{equation}
where $\mu_1$ and $u^j$ are  defined by \cite[eqs (5.14) and (2.7)]{H1}:
           \begin{equation}
           \mu_1 = -u^r \nabla _i F_r \ ^i \, ,
           \end{equation}
     \begin{equation}
              u^r={v^r \over F}\, .
           \end{equation}
Using notations of \cite{H1} throughout,  $v^r$ are fiber
coordinates of the tangent bundle over $M$ and
 $\nabla _i$ is the usual Riemannian covariant derivative defined through
 $g_{ij}(x).$
{\it Assuming} that $\mu _1 $ is
 the proper charge density \cite[p 378]{H1},
the author then identifies (1) as
the Maxwell equations {\it with source.}
The author has, therefore,
assumed that:
             \begin{equation}
                   \mu _1 =\mu_1(x) \, .
            \end{equation}
However, this {\it assumption}, together with
{\it definition} (2), already
implies equation (1).  To see this, differentiate
(2) with respect to $v^j$ and then use
(4) to obtain:
$$
\nabla_iF_j \ ^i =u^r {\partial F \over \partial v^j} \nabla _i F_r
\ ^i \, ;
$$
noting that $ {\partial F \over \partial v^j}=u_j,$ and using
 (2)  again, we
arrive at  (1).
Therefore, rather than being derived, (1) has in fact been merely assumed.

More importantly, assumption (4) implies that $\mu_1 =0$,
so that {\it the assumed
equations can be identified only as  source-free Maxwell equations.}
To see this simply differetiate
 (1) with respect to $v^k$ to obtain,
  $$ 0= \mu_1 (\delta ^j \, _k - u^j u_k)  \, . $$
  To clarify this apparently curious result, we note that in
the tangent bundle picture of Finsler geometry, fiber coordinates $v^k$, are
parameters independent of $x$. However, for a system of charged particles, for
which we can write Maxwell equations, the velocity vector is a function of
$x$.  Therefore  (1) {\it can not} be identified as
Maxwell equations {\it with source}
because $u^j$ in this equation are independent of $x$
and (contrary to \cite{H1}, p 378) cannot be considered as a velocity field.

There is, in fact, a genuine derivation of source-free Maxwell equations
within the same framework.
From the  connection given in \cite{H1} (eqs (5.1)\mbox{--}(5.3)) we can
directly calculate $H$,
the second contraction of the Berwald curvature, $H^i \,_{jkl}$,
in tems of the Riemannian scalar curvature, $R,$ and the
Maxwell tensor:
  \begin{equation}
       H=R+{3 \over 2} K^2F_{rs} F^{rs}-{5 \over 2} K \mu_1 \, .
         \end{equation}
Equation (5) is the same as eq (5.62) of \cite{H1}, however,
in derivation of the latter,
Maxwell equations have been used.
This is unnecessary and could have been easily  avoided
if, in its derivation,  eq (5.58) was used
in place of eq (5.59). Now since $H=H(x)$ from
the metric variations (see \cite{H1}, p 378),
we conclude from
(5) that $\mu _1 $ is a function of $x$ only.
Therefore, we have {\it derived} (4), and as
explained before, this easily  leads to the source-free Maxwell equations.

Comming to the treatment of
 Einstein\mbox{--}Maxwell equations in \cite{H1}, we note that,
because the mass density, $\rho_1$, has been curiosly assumed \cite[p 379]{H1}
to be proportional to $\mu_1$ \cite{note}; it, too, must be zero.
Consequently the proposed geometric formulation of Einstein\mbox{--}Maxwell
system can pertain only to source-free situations.


\begin{thebibliography}{9}
\bibitem{note}
The charge-to-mass ratio of the particles is, therefore, implicitly
assumed to be the same.
\end{thebibliography}
\end{document}